\documentclass{article}
\usepackage{mciteplus}
\usepackage{authblk}
\usepackage{lineno,hyperref}
\usepackage[squaren]{SIunits}
\usepackage{color}
\usepackage{amsmath}
\usepackage{caption,subcaption}
\usepackage{setspace}
\usepackage{multirow}
\usepackage{threeparttable}
\usepackage{booktabs}
\usepackage{graphicx}
\usepackage{dcolumn}
\usepackage{bm}
\usepackage{float}
\definecolor{americanrose}{rgb}{1.0, 0.01, 0.24}
\definecolor{coralpink}{rgb}{0.97, 0.51, 0.47}
\definecolor{ao(english)}{rgb}{0.0, 0.5, 0.0}
\definecolor{darkpastelgreen}{rgb}{0.01, 0.75, 0.24}
\definecolor{cyan(process)}{rgb}{0.0, 0.72, 0.92}
\definecolor{brown}{rgb}{0.8, 0.5, 0}

\newcommand\adenine{S4}
\newcommand\purine{S5}

\begin{document}

\captionsetup{justification   = raggedright,
              singlelinecheck = false}

\title{Vibronic fine structure in the nitrogen 1s photoelectron spectra from Franck-Condon simulations II: Indoles}

\author[1]{Minrui Wei}
\author[1]{Lu Zhang}
\author[2]{Guangjun Tian}%
\author[1]{Weijie Hua$^{\ast,}$}

 \affil[1]{MIIT Key Laboratory of Semiconductor Microstructure and Quantum Sensing, Department of Applied Physics, School of Science, Nanjing University of Science and Technology, 210094 Nanjing, China}
\affil[2]{%
 Key Laboratory for Microstructural Material Physics of Hebei Province, School of Science, Yanshan University, 066004 Qinhuangdao, China}%
\affil[ ]{$\ast$~E-mail: wjhua@njust.edu.cn (W. Hua)}

\maketitle

\begin{abstract} 
The vibronic coupling effect in nitrogen 1s X-ray photoelectron spectra (XPS) was systematically studied for a family of 17 bicyclic indole molecules by combining Franck-Condon simulations (including the Duschinsky rotation effect) and  density functional theory.  The simulated vibrationally-resolved spectra of 4 molecules agree well with available experiments. Reliable predictions for this family  further allowed us to summarize rules for spectral evolution in response to three types of common structural changes (side chain substitution,  CH$\leftrightarrow$N replacement, and isomerization). Interestingly,   vibronic properties of amine and imine nitrogen are clearly separated: they show negative and positive $\Delta$ZPE (zero-point vibration energy of the core-ionized with respect to the ground state), respectively, indicating flatter and steeper PESs induced by the N 1s ionization; amine N's show stronger mode mixing effects than imine N's; the 1s ionizations on two types of nitrogens led to distinct changes in local bond lengths and angles. The rules are useful for a basic understanding of vibronic coupling in this family, and the precise spectra are useful for future reference and data mining studies.
  \end{abstract}


\section{Introduction} 
X-ray photoelectron spectroscopy (XPS) is widely used to characterize molecular and material structures nowadays. The core-level binding energy (BE) is both element-selective and sensitive to local bonding. For gas molecules, high-resolution vibrationally-resolved XPS spectra provide richer information beyond binding energies: the fingerprint signatures reflect information about the potential energy surfaces (PESs) of the initial (ground) and final (core-ionized) states, as well as the electronic and nuclear dynamics during the core ionization process.\cite{hergenhahn_vibrational_2004, mendolicchio_theory_2019, plekan_investigation_2022, plekan_theoretical_2008, myrseth_vibrational_2002, fronzoni_vibrationally_2014, sankari_vibrationally_2003} 

The groundbreaking work of Kai Sieghabn\cite{siegbahn1982electron, siegbahn1967atomic, siegbahn1974esca} since the 1950s has significantly improved the precision of XPS [i.e., electronic spectroscopy for chemical analysis (ESCA) as coined by him]. Since then, a vast amount of XPS data has been collected over the couple of decades. The technique has enabled the characterization of the structures of lots of molecules and materials and helped scientists  understand molecular physics, guide the material design, and so on. In such a characterization process, spectral interpretation is an essential final step that translates the detected numerical data into physical and chemical insights.  Various experimental databases\cite{Rumble1992NIST, wagner_1979_handbook, jolly_core-electron_1984, xpsdatabase, sasj, LaSurface} were constructed to assist the interpretation process, with the most famous being the National Institute of Standards and Technology (NIST) database.\cite{Rumble1992NIST} 

However, existing experimental XPS databases have two major drawbacks: variations in BEs among different experiments and the lack of profile data.  Different experiments of the same compound reported discrepant BEs, sometimes over 1 eV.\cite{wei_vibronic_2022, Rumble1992NIST} The discrepancy mainly  comes from the calibration process which can be highly arbitrary.\cite{greczynski_x-ray_2020} Meanwhile, most databases collect only BE values, without including the spectral profiles. Efforts were also made to build online databases,\cite{xpsdatabase, sasj, LaSurface} where raw pictures of the original experimental spectra, with both the BE and profile information,  were collected, which facilitated the comparison process for various different compounds.  However, the available information is still limited, and possible inconsistency in different experiments can impede close analyses. 

For these two reasons, constructing a theoretical library for high-resolution XPS is necessary. Data achieved on the same footing provides fair accuracy to analyze the physical rules behind it, especially among structurally-similar systems. Before generating huge data, at the current stage, it is more important to guarantee reliable data. We wish to validate and optimize the simulation procedure and try to investigate general rules for vibronic coupling based on limited molecules with specific structural similarities. Our initial studies have shown great agreement with the experiments by combing the full core hole (FCH) density functional theory (DFT) and the Franck-Condon simulations.\cite{hua_theoretical_2020, wei_vibronic_2022, cheng_vibrationally-resolved_2022} This motivates us to expand the use of this protocol to wider systems and gain more comprehensive insights into vibronic coupling within a family. Quantitative assessment of the impact of structural changes on molecular properties is meaningful for further understanding the structure-spectroscopy relation, which is insightful for future data mining studies.

Aromatic nitrogen-containing heterocyclic molecules (N-heterocycles)\cite{joule2020heterocyclic} serve as an ideal family for testing new theoretical methods,\cite{damour_accurate_2021, *da_silva_retro-3_2008, *gutowski_accurate_2006, *verevkin_rediscovering_2012} which are also important building blocks for biomolecules,\cite{kerru_review_2020} high energy-density compounds,\cite{zheng_self-assembly_2021} nonlinear optical materials,\cite{castro_design_2012} and pharmacologically active compounds.\cite{mermer_recent_2021} In part I\cite{wei_vibronic_2022} of this series of papers, our calculations on pyrimidine have shown good agreement with the high-resolution gas-phase experiment,\cite{bolognesi_pyrimidine_2010}  accurate vibrationally-resolved N1s XPS spectra  for a group of azine molecules were predicted and analyzed to understand the rules for vibronic coupling effects as influenced by consecutive replacement of the CH group with an N atom.

As a continuation, in this work (paper II) we further investigate the indole family.  Indole has long been recognized as a star molecule in the field of biochemistry, being a structural motif for numerous drugs and natural products.\cite{sundberg_2012_chemistry, gribble_2016_indole, jia_current_2020, dorababu_indole_2020, han_importance_2020} Structurally, azines are six-membered rings, indoles are fused by one six- and one five-membered rings.  Besides, we also plan to present our predictions for five-membered ring compounds (where there are less experimental data) in a future separate work. We hope the three papers can construct a complete picture to understand the N1s vibronic fine structure for small aromatic N-heterocycles. 

\begin{figure*}
\includegraphics[width=0.8\textwidth]{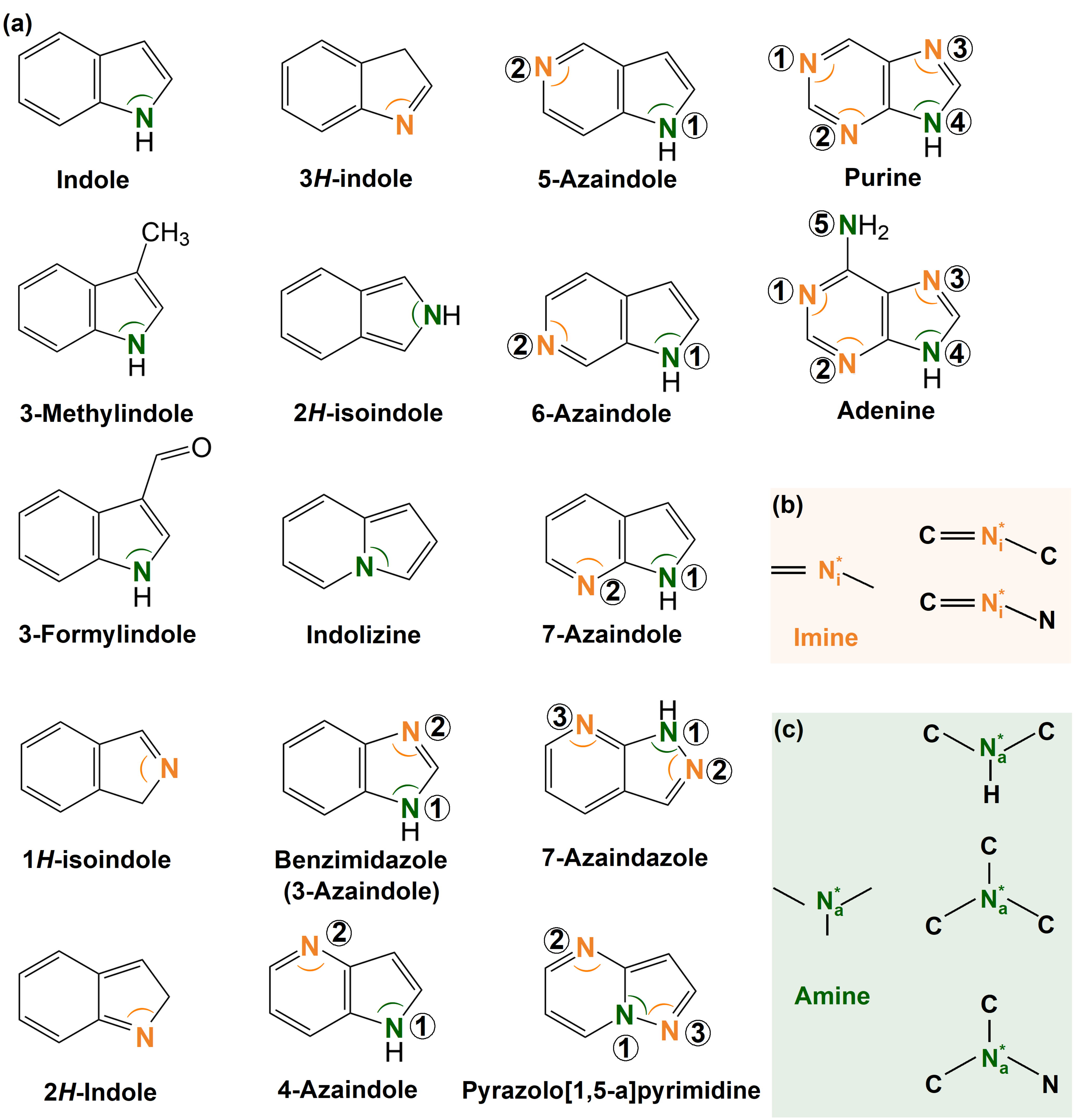}
\caption{(a) Structures of all 17 molecules under study, aligned in  column-major order with increasing (1, 2,$\cdots$, 5) N atoms. In molecules with more than one N atom, all N atoms are indexed by numbers in circles (not to be confused with the atomic indices as used in nomenclature). Amine (green) and imine (orange) N's are distinguished by colors. Selected angles are labeled by arcs. (b) Two imine  and (c) three amine  type local structures at the ionized N center (denoted by N$_\text{i}^*$ and N$_\text{a}^*$, respectively) extracted from the 17 molecules. 
}
\label{fenzi}
\end{figure*}

Figure \ref{fenzi}(a) depicts the  17 selected common indole-derived molecules. Our choice of sample set covers three common types of structural changes: (1) isomerization, (2) side chain substitution, and (3) CH$\leftrightarrow$N replacement. We chose 6 indole isomers in this study, where three contain imine (=N--) and three contain amine (--N$<$) nitrogens [Fig. \ref{fenzi}(b,c)]. This sample set is to examine the structure-spectroscopy relation in response to the local bonding change at the ionization center N$^*$ (N$_\text{i}^*$ or N$_\text{a}^*$ for imine or amine nitrogen). Significant differences between the two types were found for N1s binding energies,  X-ray absorption (XAS) spectra, and resonant inelastic X-ray scattering (RIXS) spectra of small molecules,\cite{du_theoretical_2022, hua_systematic_2010} large DNA double strands,\cite{hua_systematic_2010} and  two-dimensional material g-C$_3$N$_4$.\cite{zhang_accurate_2019} We will investigate the XPS fine structures and vibronic properties as related to the two types. 

Side chain substitution can push the electrons to or pull the electrons off the indole ring. Two substituted derivatives with available experimental spectra were selected in our study: one -CH$_3$ (3-methylindole) and one -CHO (3-formylindole) substitute.  -CH$_3$ is a weak electron-donating group (EDG) and -CHO is a moderate electron-withdrawing group (EWG). The two molecules are important precursors for the synthesis of some biologically or pharmacologically active compounds.\cite{liljefors_2002_textbook, *ebrahimi_new_2013}         

Meanwhile, 9 molecules were chosen to demonstrate the different degrees (1, 2, 3) of CH$\leftrightarrow$N replacement. Various azaindoles represent an important class of organic molecules with one N substitution on the indole ring, which are fine chemical intermediates commonly used in medicine,\cite{ meanwell_inhibitors_2018} biological materials,\cite{lee_azaindolylsulfonamides_2014} and natural products.\cite{walker_variolins_2009}  A total of 5 azaindoles were selected, among which benzimidazole (also named 3-azaindole) is a special one where two nitrogens both locate in the five-membered ring (cf. the rest, one nitrogen on one ring). Besides, 2 molecules each for double (7-azaindazole and pyrazolo[1,5-a]pyrimidine)  and triple (purine and adenine) CH$\leftrightarrow$N replacements were selected. Structurally, an adenine is simply an -NH$_2$ substituted purine. Adenine is also one of the building block molecules for DNA, responsible for the radiation loss and photolysis properties. 

Experimental N1s XPS spectra, to our knowledge, are only available for indole,\cite{plekan_experimental_2020} 3-methylindole,\cite{zhang_electronic_2009} 3-formylindole,\cite{plekan_experimental_2020} and adenine\cite{plekan_theoretical_2008} among all systems under study. The spectral resolutions are not always high enough to see clear vibronic structures (especially for indole and 3-formylindole\cite{plekan_experimental_2020}), but the peak asymmetry is obvious for every molecule, indicating significant influences of the vibronic coupling effects. On the other hand, theoretical XPS studies were done for selected systems, but limited to pure vertical excitations and electronic-only calculations. For example, recently, He et al.\cite{he_2022_specific}  simulated the N1s XPS spectra of indole at the MP2 level with a relatively large half-width-at-half-maximum (hwhm) of 0.58 eV and achieved a generally similar profile to the experiment. Plekan et al.\cite{plekan_theoretical_2008} and Wang et al.\cite{wang_inner-shell_2008} simulated the XPS spectra of adenine by DFT with relatively large shifts in absolute binding energies (-1.32\cite{plekan_theoretical_2008} and 1.65 eV\cite{wang_inner-shell_2008}). The goal of this study is to provide high-precision vibrationally-resolved XPS spectra, based on which to carry out systematic analyses (dominant vibronic transitions, active vibrational modes, contributions of 0-$n$ transitions, structural changes induced by core ionization, mode mixing, etc.) and yield general rules on the vibronic coupling.

 \section{Computational methods} \label{sec:method}
Details were presented in paper I.\cite{wei_vibronic_2022} Briefly, all electronic structure calculations were first performed  by using  Gamess-US,\cite{schmidt_general_1993, gordon_advances_2005} where the DFT method with the B3LYP functional\cite{becke_density-functional_1988, *becke_new_1993, *lee_development_1988} was employed. Then, Franck-Condon simulations with the inclusion of the Duschinsky rotation (DR)\cite{duschinsky_1937} effects were used to generate the vibronic fine structures by using the modified\cite{hua_theoretical_2020} DynaVib package.\cite{DynaVib} All ground state (GS) calculations were performed by using the cc-PVTZ basis set.\cite{dunning_gaussian_1989, kendall_electron_1992} A consistent basis set was adopted for the core-ionized state and
a double basis set technique\cite{hua_theoretical_2020, wei_vibronic_2022} was used. Vertical and adiabatic ionic potentials (IPs), and the 0-0 vibrational transition energy were computed respectively via\cite{hua_theoretical_2020, wei_vibronic_2022} 
\begin{eqnarray}
I^\text{vert} &=& 
E_\text{FCH}|_\mathbf{min\, GS} -  E_\text{GS}|_\mathbf{min\, GS} + \delta_\text{rel},
\label{eq:IP:vt}\\
 I^\text{ad}  &=&  E_\text{FCH}|_\mathbf{min\,FCH} - E_\text{GS}|_\mathbf{min\, GS} + \delta_\text{rel}, \label{eq:IP:ad} \\
 E_{00}^\text{DR}   &=&  I^\text{ad} + \Delta {\varepsilon}_{0}.\label{eq:E00:DR}
\end{eqnarray}
Here  $E_\text{GS}$ ($E_\text{FCH}$) denotes the total energy of the GS (FCH) state, and  \textbf{min GS} (\textbf{min FCH}) represents the optimized geometry of the GS (FCH) state. $\Delta {\varepsilon}_{0}$ is the difference  of the zero-point vibrational energies (ZPE) between the FCH ($\varepsilon_0^\text{FCH}$) and GS ($\varepsilon_0^\text{GS}$) states, i.e.,
\begin{equation}
\Delta {\varepsilon}_{0} = \varepsilon_0^\text{FCH} -  \varepsilon_0^\text{GS}.
\end{equation}

Stick spectra were convoluted by a Lorentzian line shape and slightly different hwhm's were set to better compare with different experiments:\cite{plekan_experimental_2020, zhang_electronic_2009, plekan_theoretical_2008} 0.03 eV, indole, 3-methylindole, and 3-formylindole; 0.05 eV, others.  For each molecule with multiple nitrogens, its total spectrum was calculated simply by summing the individual atom-specific contributions.  Besides, major stick transitions were analyzed, where thresholds for FCFs, F$\geq$0.02 (indole, 3-methylindole, and 3-formylindole) or F$\geq$0.04 (the rest) were used to filter out weak transitions.  All major assignments of each individual molecule are provided in the Supplemental Material\cite{si_indole} (Figs. S1--S17; emphasized with colored vertical lines). For those with available experiments (indole,\cite{plekan_experimental_2020} 3-methylindole,\cite{zhang_electronic_2009} 3-formylindole,\cite{plekan_experimental_2020} and adenine\cite{plekan_theoretical_2008}), theoretical spectra were shifted by 0.25, 0.32, 0.22, and 0.44 eV, respectively, to better compare with the experiments. No \emph{ad hoc} shift was applied for the rest.

\section{\label{sec:level3} Results}
\subsection{Statistics on vertical and adiabatic IPs}
\begin{figure*}
\includegraphics[width=0.8\textwidth]{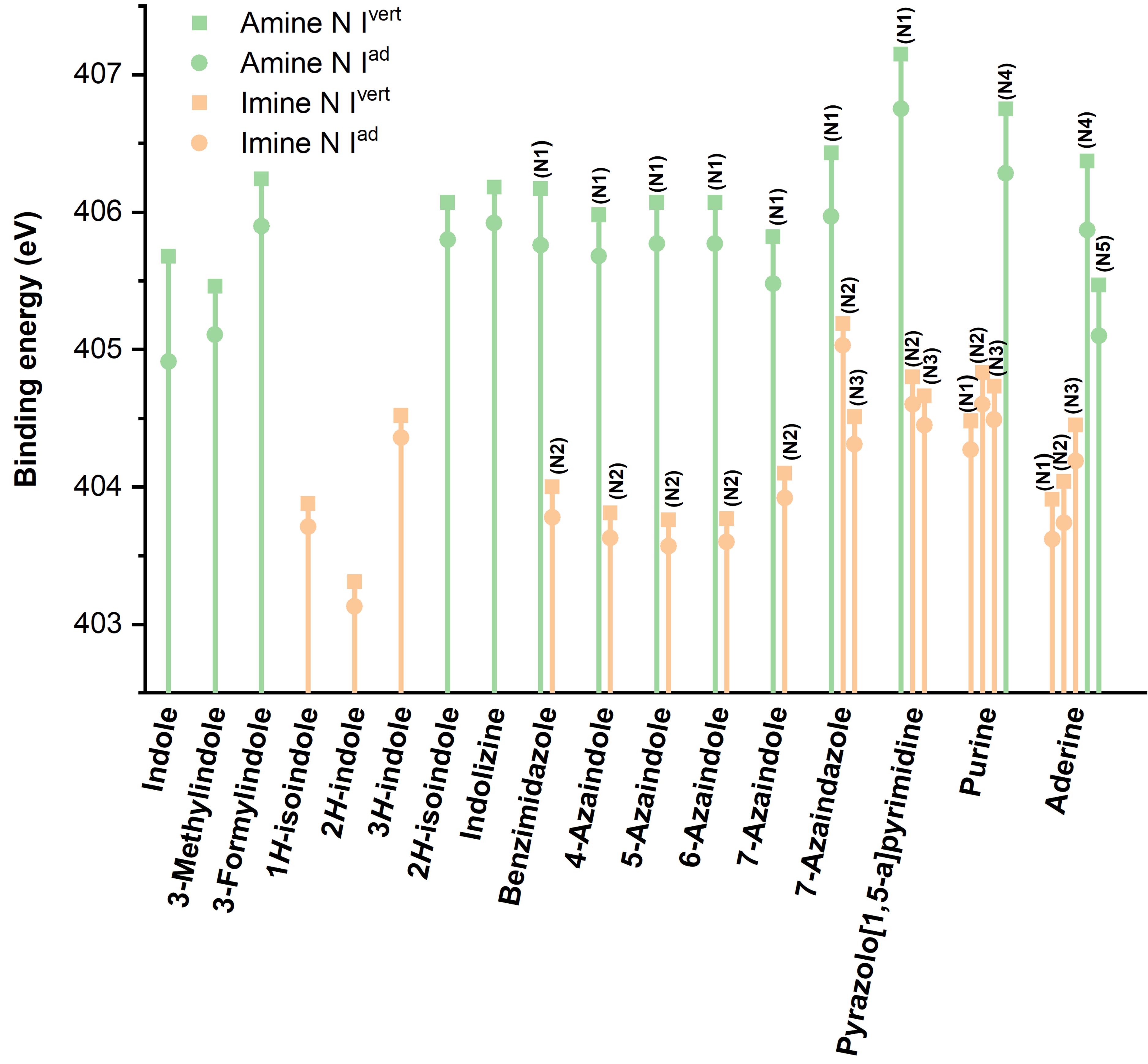}\caption{Simulated vertical ($I^\text{vert}$, square) and adiabatic ($I^\text{ad}$, circle) ionization potentials of all molecules.  Amine (green) and imine (orange) nitrogens are distinguished in color.  See definition of each nitrogen in Fig. \ref{fenzi}(a).}
\label{ip-tol}
\end{figure*}

Table \ref{tab:ip} lists the theoretical vertical and adiabatic N1s IPs for the 33 nitrogen sites of all 17 molecules. The data is further visualized in Fig. \ref{ip-tol}, where a clear separation of 15 amine and 18 imine nitrogens is vividly illustrated.  All theoretical vertical IPs are in a range of 403.3--407.2 eV spanning over 3.9 eV. All amine N's locate at higher (405.5--407.2 eV) and all imine N's at lower (403.3--405.2 eV) energy regions, with a gap of ca. 0.3 eV. Comparisons were made with available experiments\cite{plekan_experimental_2020, zhang_electronic_2009, plekan_theoretical_2008} for selected molecules, and the deviations of vertical IPs are from -0.4 to +0.1 eV. The deviations are consistent with those found in $\Delta$Kohn-Sham calculations of other molecules.\cite{bagus_consequences_2016, pueyo_bellafont_validation_2015, pueyo_bellafont_prediction_2015, pueyo_SCF_performance_2016, du_theoretical_2022} 

We also compared the adiabatic IPs with experiments,\cite{plekan_experimental_2020, zhang_electronic_2009, plekan_theoretical_2008} which show larger deviations ({-0.8} to {-0.2} eV) than vertical calculations. Adiabatic IPs cover a range of 3.7 eV (403.1--406.8 eV), with the amine  and imine N's being in the higher- (405.1--406.8 eV) and lower-energy (403.1--405.0 eV) regions, respectively. Overlap of the two regions appears owing to the vibronic coupling effects.

\subsection{Statistics on $\Delta I$ }

For each N1s ionization, the adiabatic IP is about 0.2--0.5 eV lower than the corresponding vertical IP. By combining Eqs. (\ref{eq:IP:vt})-(\ref{eq:IP:ad}),  their  difference can be computed by the total energy change in the FCH-state PES:
\begin{equation}
\Delta I \equiv I^\text{vert} - I^\text{ad} = 
E_\text{FCH}|_\mathbf{min\,GS} - E_\text{FCH}|_\mathbf{min\,FCH}.
\end{equation}
Here $\Delta I$  describes the structural relaxation effect  in the excited-state PES caused by the  core ionization. This parameter approximately (neglecting the curvature change) provides an estimate of the displacement in PESs. From our calculations, it is interesting to find that  $\Delta I$  for amine N's (0.3--0.5 eV) are generally larger than imine N's (0.2--0.3 eV).  We deduce that displacements in PESs of amine N's are generally larger  than imine N's. 

\subsection{Statistics on structural changes} 

\begin{figure*}
\includegraphics[width=0.8\textwidth]{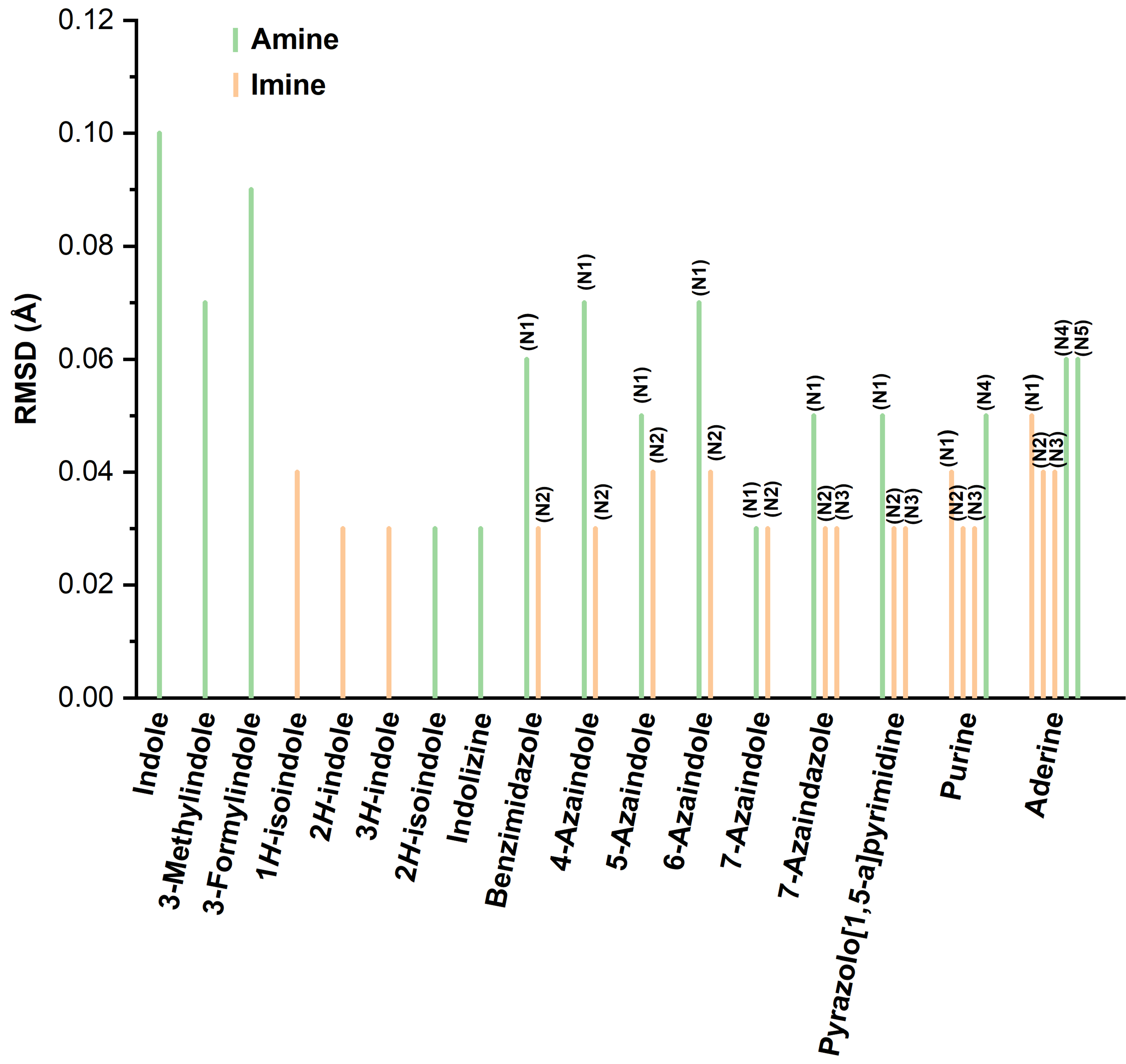}
\caption{RMSD values between two Cartesian coordinates, the optimized core-ionized (\textbf{min FCH}) and the ground (\textbf{min GS}) states, for each N1s ionization of each molecule. Amine (green) and imine (orange) nitrogens are distinguished in color. See definition of each nitrogen in Fig. \ref{fenzi}(a).}
\label{rmsd}
\end{figure*}

This deduction is supported by Fig. \ref{rmsd}, where root-mean-squared distances (RMSDs) between the ground and excited state structures are plotted for each nitrogen ionization. It can be seen that ionizations of amine N's  lead to relatively larger RMSDs (0.03--0.10 \AA) than imine N's  (0.03--0.05 \AA). This can be understood that an N=C double bond (as in imines) can be more ``rigid'' than two N--C (or N--N) single bonds (as in amines) during the N1s ionization process.

\begin{figure*}
\includegraphics[width=1.0\textwidth]{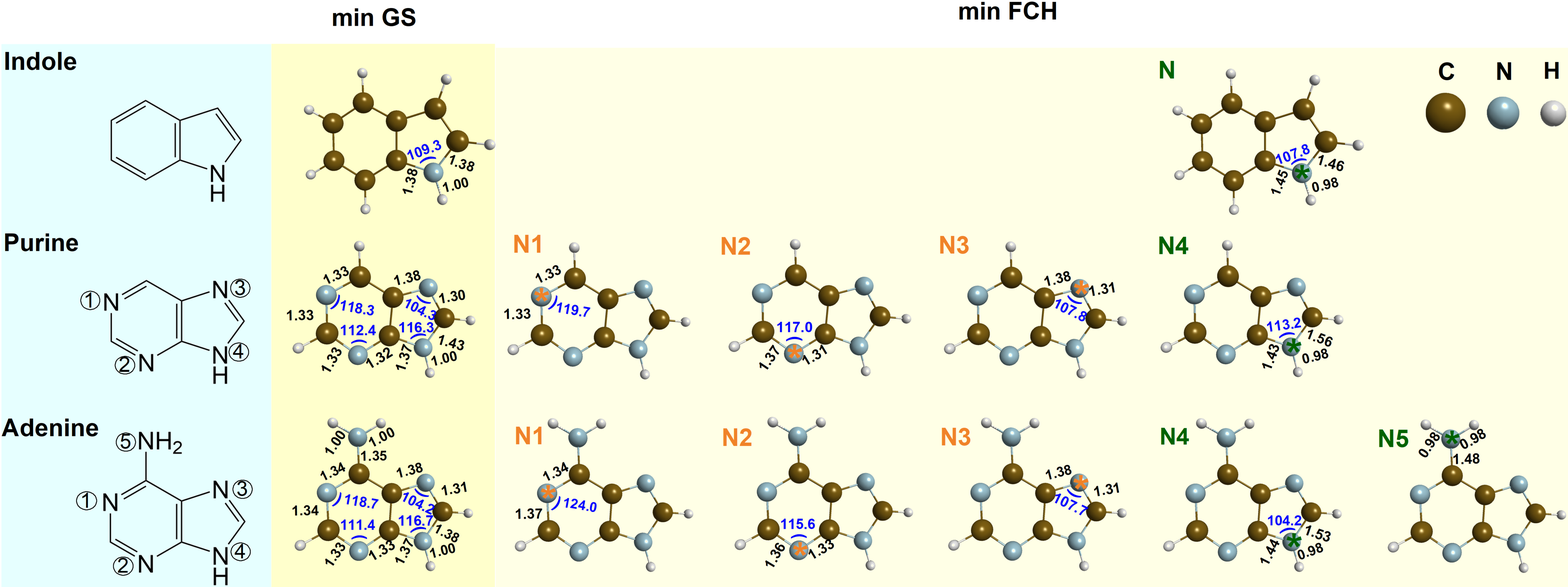}
\caption{Optimized geometries of three selected molecules in the ground (\textbf{min GS}) and core-ionized (\textbf{min FCH}) states, where selected bond lengths (in \AA) and angles (in $^\circ$) at N$^*$ are labeled. Corresponding Kekul\'{e} structures are also recaptured [from Fig. \ref{fenzi}(a)] on the left for comparisons.}
\label{jiegou}
\end{figure*}

Figure \ref{jiegou} compares the optimized structures of the ground and the core-ionized states for three selected systems indole, purine, and adenine.  It is worth noting that the left of Fig. \ref{jiegou} (and also Fig. \ref{fenzi}) illustrates only one of the Kekul\'{e} structures for each molecule. A conjugated $\pi$ bond system and resonant Kekul\'{e} structures average away the difference of those single and double bonds (especially in the six-membered ring) in the geometrical optimizations. We thus simply use N--C (N$^*$--C) to denote the bond length in any range between the two atoms at \textbf{min GS} (\textbf{min FCH}). For instance, for N2 in adenine (Fig. \ref{jiegou}), the two N2--C distances at \textbf{min GS} are equal (1.33 {\AA}). The equality is broken by core ionization, and the resulting N2$^*$-C bond lengths become 1.36 and 1.33 {\AA}, respectively. Similarly in purine, the two almost equal bond lengths (1.33 and 1.32 {\AA}) in the ground state become 1.37 and 1.31 {\AA} respectively in the core-ionized state.  The structural changes induced by N1 ionization in adenine and purine are different. In purine, the two N1--C bond lengths (both 1.33 {\AA}) stay almost unchanged after the ionization. While in adenine, the two N1--C lengths (both 1.34 {\AA}) become 1.37 and 1.34 {\AA}, respectively after the ionization. This is  because the -NH$_2$ substitution in adenine is close to N1 and changes its local environment, which leads to different 
structural responses in both molecules to the 1s ionizations.

In the five-membered ring, the two N-C distances at each N site seem to better agree with the Kekul\'{e} structures, showing evident one long and one short distance. At N3 (see Fig. \ref{jiegou}), the two N-C distances are respectively 1.38 and 1.30 {\AA} in purine (1.38 and 1.31 {\AA} in adenine), which are almost unchanged after the N1s ionization. While evident elongation happens at N4, which is 0.07 and 0.08 {\AA} in indole, 0.06 and 0.13 {\AA} in purine, and 0.07 and 0.15 {\AA} in adenine. For the three molecules, all N-H distances are slightly reduced by 0.02 {\AA} from 1.00 to 0.98 {\AA}.

Table \ref{tab:st} presents more complete data for bond lengths and angles at N$^*$ of all the 17 molecules at \textbf{min FCH}.  The N$_\text{a}^*$-H bond lengths stay almost a constant value of 0.98 {\AA}, decreasing by only 0.02 or 0.03 {\AA} from the GS geometry.  For the N$^*$--X distances [X=C (mainly), N (rarely)], amine nitrogens (N$_\text{a}^*$-X, 1.40--1.56 {\AA}) always show longer values than imine nitrogens (N$_\text{i}^*$-X, 1.30--1.43 {\AA}).  The change of N$^*$--X length as compared to the GS geometry is different for amine and imine N's: N$_\text{a}^*$-X is always elongated by  {0.04-0.18}{\AA}; while N$_\text{i}^*$-X can be either elongated or shortened by 0.00-0.03 {\AA}. Concerning the bond angles $\angle$C-N$^*$-X, we found a decrease for all amine N's ($\angle$C-N$_\text{a}^*$-X) and an increase for all imine N's ($\angle$C-N$_\text{i}^*$-X). 

\subsection{Statistics on ZPE changes ($\Delta {\varepsilon}_{0}$)}

\begin{figure*}
\includegraphics[width=0.8\textwidth]{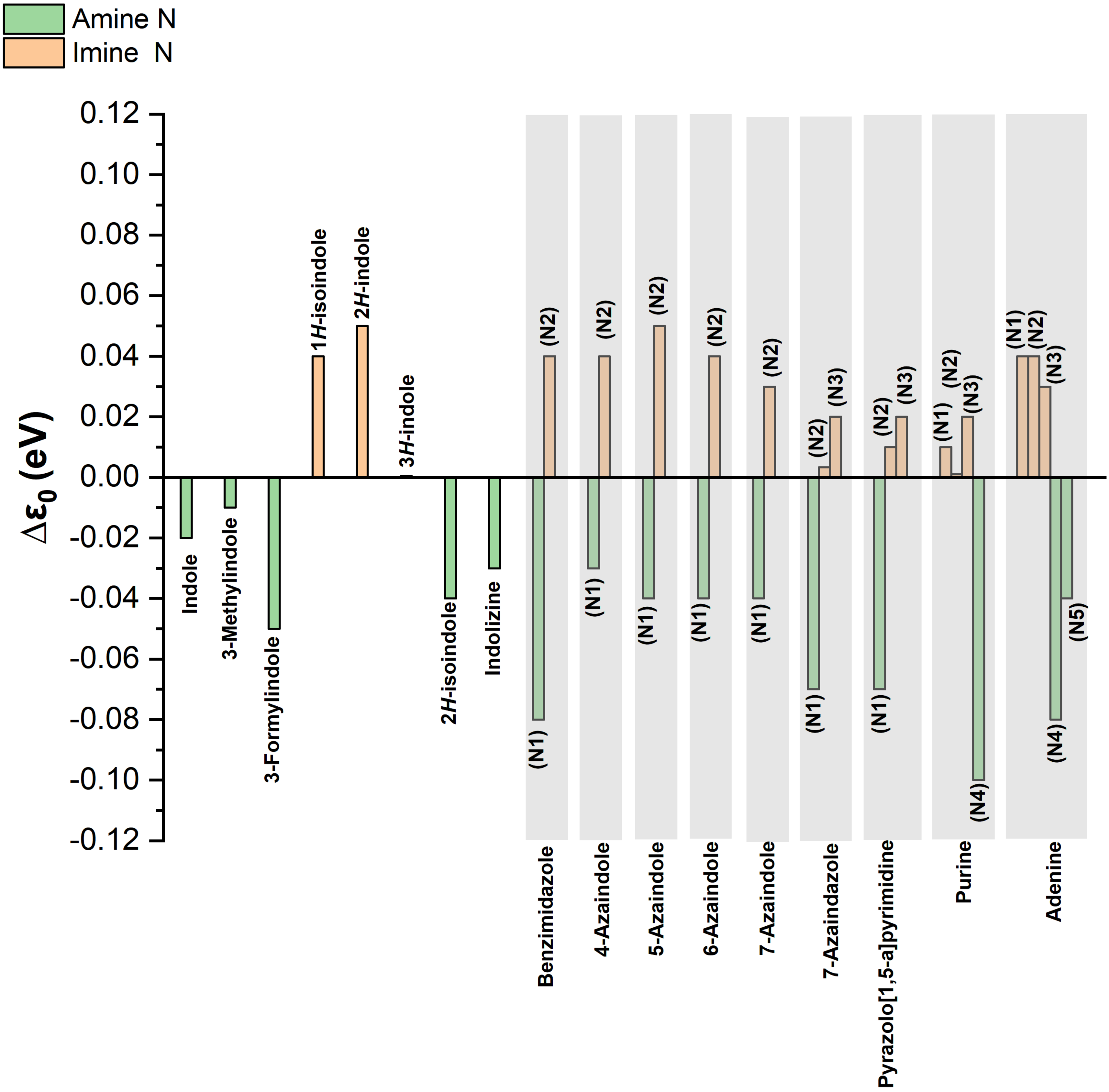}
\caption{The  $\Delta {\varepsilon}_{0}$ values (difference of ZPE in the final state as compared to that in the initial state) of amine and imine nitrogens in all molecules.  $\Delta {\varepsilon}_{0}$ of amine (green) and imine (orange) nitrogens are distinguished by colors.  See definition of each nitrogen in Fig. \ref{fenzi}(a).}
\label{zpe}
\end{figure*}

To estimate the deformation of the excited-state PESs from the ground-state ones as induced by core ionization, we analyzed the $\Delta {\varepsilon}_{0}$ values (the zero-point vibrational energies in the core-ionized as referred to the ground states) of all these 33 nonequivalent  N sites.  As depicted in Fig. \ref{zpe}, we found that amine (imine) nitrogens show positive (negative) $\Delta {\varepsilon}_{0}$ values, which are about -0.1 to {-0.01} (0.00 to +0.05) eV.  Our results indicate that N1s ionization leads to a distinct change direction in the curvature of the final-state PES, which becomes flatter (steeper) for amine (imine) N's.

\section{\label{sec:level4} Discussion}

\subsection{Effects of side chain substitutions}
\subsubsection{-CH$_3$ and -CHO substitutions on indole} 
\begin{figure*}
\includegraphics[width=0.9\textwidth]{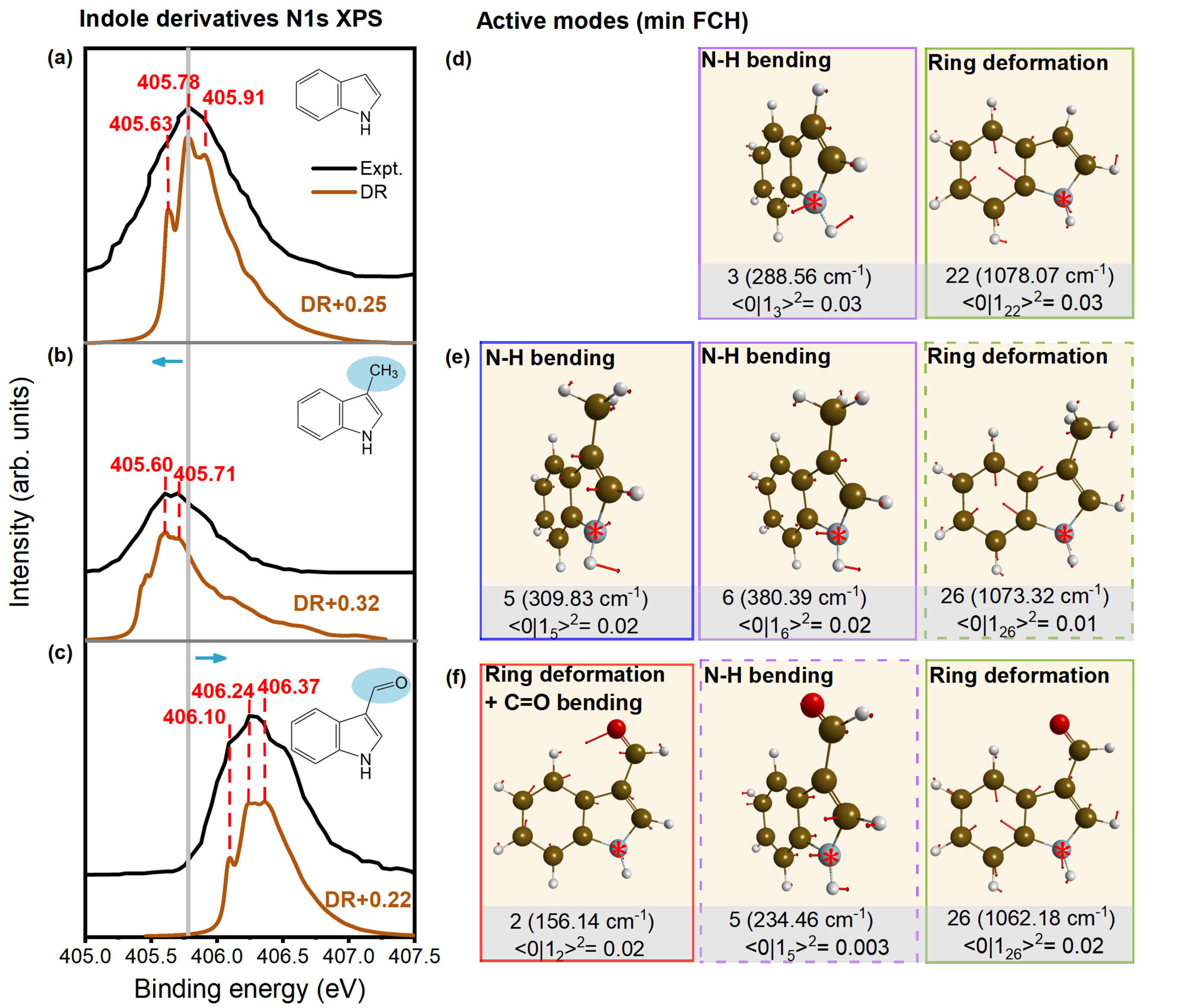}
\caption{Simulated vibrationally-resolved  N1s XPS spectra  of (a) indole, (b) 3-methylindole, and (c) 3-formylindole  by using the FCH-DR method.  Theoretical spectra are uniformly shifted by $\delta$=0.25, 0.32, and 0.22 eV to better compare with experiments (Plekan et al.,\cite{plekan_experimental_2020} Zhang et al.,\cite{zhang_electronic_2009} and Plekan et al.\cite{plekan_experimental_2020} for the three molecules, respectively).  (d-f)  Active vibration modes of the three molecules at the final state structure (\textbf{min FCH}) (in solid frames, colored to show different mode types). Additionally, two less active modes are shown for comparison in dashed frames. Stars indicate N$^*$. See text for details.} 
\label{3indole}
\end{figure*}

 Figure \ref{3indole}(a-c) depicts our computed vibrationally-resolved N1s XPS spectra of indole, 3-methylindole, and 3-formylindole compared to   experiments.\cite{plekan_experimental_2020, zhang_electronic_2009}  Spectra of the three molecules exhibit significant differences. The (small, ca. 0.2 eV) red and (moderate, ca. 0.4 eV) blue shifts in 3-methylindole  and 3-formylindole are simply because  -CH$_3$ and -CHO are a (weak) electron-donating and a (moderate) electron-withdrawing groups, respectively. 
 
 The experimental spectrum of indole\cite{plekan_experimental_2020} shows only an asymmetric broad peak at 405.8 eV. Our theoretical spectrum well reproduced the general profile and further identifies three characteristic peaks at 405.6, 405.8, and 405.9 eV, which arise from (0-0, 0-1), (0-1, 0-2), and 0-2 transitions, respectively [Fig. S1(b)]. The experimental spectrum of 3-methylindole\cite{zhang_electronic_2009} resolved two small peaks at 405.6 and 405.7 eV, with a separation of 0.1 eV. Our simulation well reproduced the fine structures, which were interpreted as 0-2 and 0-3 transitions, respectively [Fig. S2(b)].  Concerning 3-formylindole, 0-0 and 0-1 transitions contribute a feature at 406.1 eV, and 0-1, 0-2, and 0-3 transitions contributed to the high-energy region at 406.2--406.4 eV [Fig. S3(b)].

Figure \ref{3indole}(d-f) illustrates the active vibrational modes (in solid frames) of each molecule identified according to the threshold of {F$\geq$0.02}.  Two  active vibrational modes were identified for each molecule [see also Figs. S1(c)-S3(c)].  For indole (panel d), the two modes are an N-H bending mode $\nu_3$ (288.6 cm$^{-1}$) and a ring-deformation mode $\nu_{22}$ (1078.1 cm$^{-1}$).  Both modes still can be tracked in the other two substituents, but they do not always hold to be the most active modes.  For instance,  $\nu_{26}$ of the -CH$3$ substituent (panel e) and $\nu_{5}$ of the -CHO substituent (panel f) give FCFs of only 0.01 and 0.003, respectively (less than our threshold).  Meanwhile, new active modes appear in the two substituents (compared to those in indole), including another type of N-H bending mode $\nu_{5}$ (309.8 cm$^{-1}$) in the -CH$3$ substituent (panel e) and a combined ring deformation and C=O bending mode $\nu_{2}$ (156.1 cm$^{-1}$) in the -CHO substituent (panel f), both with FCFs of 0.02.  The results show that side chain substitution can effectively change the active modes, providing give sensitive isomer-dependent signatures.

Despite the difference in active modes,  the local bond lengths/angles (at N$^*$) of the three molecules in their FCH states are very similar (Table \ref{tab:st}). The N$^*$-C lengths are  between 1.45--1.47 {\AA}.  The angles $\angle$C-N$^*$-C fall between 107.3--108.0$^\circ$.  

\subsubsection{-NH$_2$ substitution on purine}

\begin{figure*}
\includegraphics[width=0.8\textwidth]{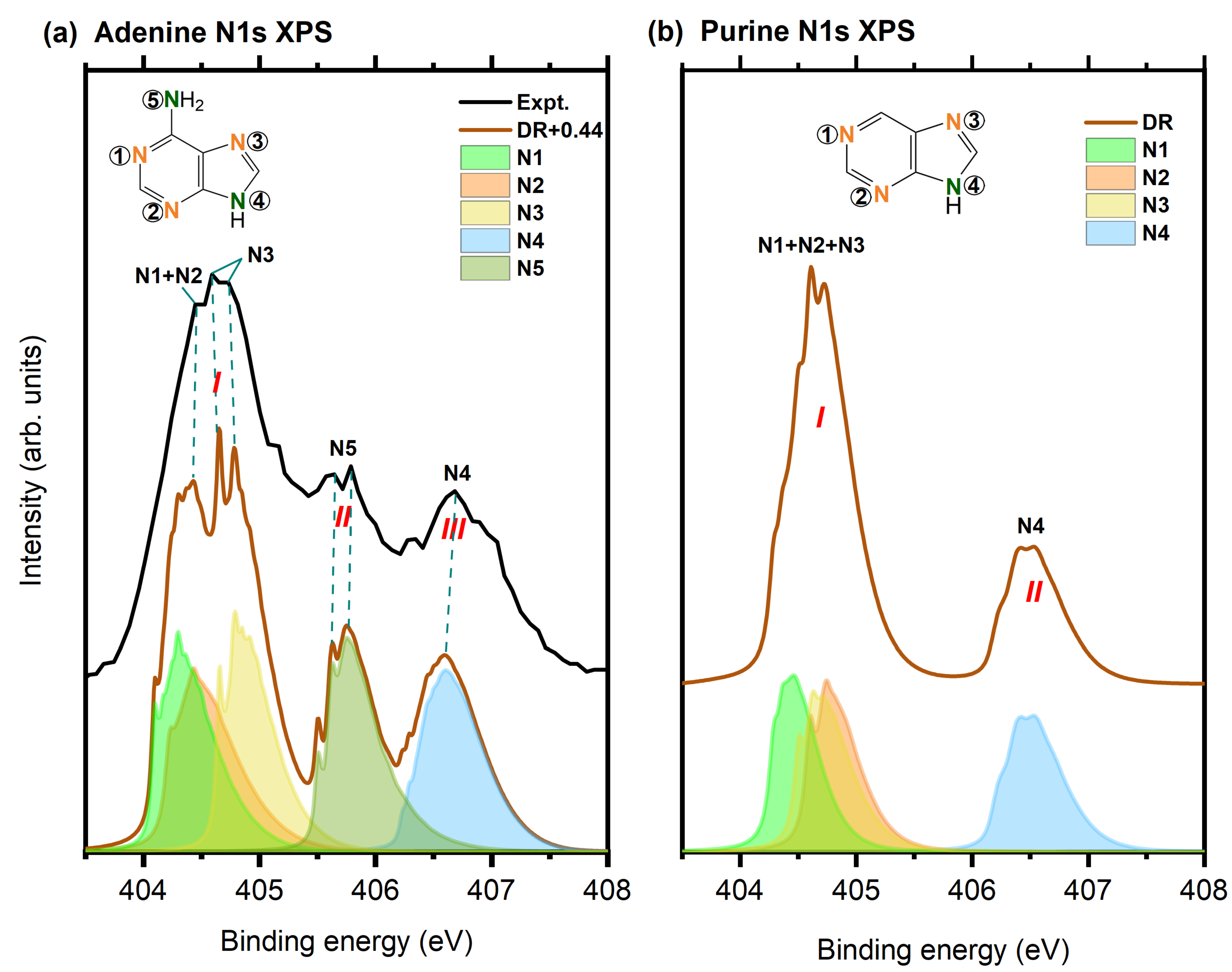}
\caption{Simulated vibrationally-resolved  N1s XPS spectrum of (a) adenine and (b) purine. Regions are labeled by Roman letters ($I$, imine N; $II$, $III$, amine N), and spectral fine structures are interpreted. Shaded areas are atom-specific contributions (see definitions in each inset).  In panel (a), the theoretical spectrum is shifted by +0.44 eV for better comparison with the experiment.\cite{plekan_theoretical_2008}
}
\label{ade}
\end{figure*}

Figure \ref{ade}(a) displays the simulated vibrationally-resolved XPS spectrum of adenine compared with the experiment.\cite{plekan_theoretical_2008}  The experimental spectrum featured three regions $I$--$III$ centered at ca. 404.4, 405.7, and 406.7 eV, respectively.  Our simulation agrees well with the experimental fine structure.  We interpreted region $I$ by imine nitrogens N1, N2, and N3, and regions $II$ and $III$ by amine nitrogens N5 (0-1, 0-2 transitions) and N4 (0-3 transition), respectively [Fig. {\adenine}(b-f)]. In region $I$, three vibronic structures were reproduced. The lowest fine structure (404.5 eV) comes from the 0-1 and 0-2 contributions of N1 and N2 [Fig. {\adenine}(b,c)], and the next two (404.6 and 404.7 eV) being respectively the 0-0 and 0-1 transitions of N3 [Fig. {\adenine}(d)]. According to our threshold (F$\ge$0.04), only N3 contains stick vibronic transitions with large  intensities [Fig. {\adenine}(a)]. Two active modes  $\nu_\text{18}$ (965.4 cm$^{-1}$) and $\nu_\text{22}$ (1164.4 cm$^{-1}$) are identified, which are both ring deformation modes localized mainly in the five-membered ring [Fig. {\adenine}(g)].

Figure \ref{ade}(b) shows the simulated spectrum of purine. Similar to adenine, imine (N1--N3) and amine (N4) nitrogens contribute to the two regions $I$ (ca. 404.6 eV) and  $II$ (ca. 406.5 eV), respectively. In region $I$, the three fine structures at 404.5, 404.6, and 404.7 eV come from mixed vibronic contributions by these three atoms [see also Fig. {\purine}(a)]. In region $II$, the doublet fine structures with nearly equal intensities (at 406.4 and 406.5 eV) are respectively assigned as 0-2 and 0-3 as well as 0-3 and 0-4 transitions of N4 [Fig. {\purine}(e)]. For each nitrogen, some 1-3 active modes are identified, which are all low-frequency (330.2--1075.8 cm$^{-1}$), in-plane ring deformation modes [Fig. {\purine} (f)].

Purine and adenine differ by an -NH$_2$ group. -NH$_2$ is a strong EDG, which leads to smaller (0.3--0.8 eV) vertical IP values for N1--N4 in adenine than in purine (Table \ref{tab:ip}). Our calculations show that in the FCH state, the bond lengths at each of the same N$^*$ sites are similar (Table \ref{tab:st}). The N$^*$-C lengths in each molecule deviate by 0.01-0.03 {\AA} compared to the corresponding GS geometry.   The differences in bond angles ($\angle$C-N$_\text{a}^*$-C and $\angle$C-N$_\text{i}^*$-C) are 0.1--4.3$^\circ$.

\subsection{Effects of isomerization}\label{sec:isomer:indole}

Figure \ref{1n}(a) depicts computed N1s XPS spectra of six indole isomers.  Above we have discussed the separation of binding energies for imine and amine N's. Concerning the vibronic profiles, they also distinguish well from each other. For each isomer, the sum of FCFs converges (a threshold of 0.99 was used) at the same $n$ = 7 [Figs. S1(b), S11(b)--S15(b)]. The isomer-dependent signatures come from different modulations of each 0-$n$ transition. For each isomer, a lower-energy 0-0 peak always exists, which is well separated from the broad peak in the higher-energy part. The broad peak is assigned to be mainly 0-1 transitions (3\textit{H}-indole) or a combination of 0-1 and 0-2 transitions (indole, 1\textit{H}-isoindole, 2\textit{H}-indole, 2\textit{H}-isoindole, and indolizine). Decomposition analyses also show that for each molecule, 0-1 transitions always make the largest contribution among all 0-$n$ transitions. This indicates a relatively small displacement of the PESs as induced by core ionization, owing to the stiffness of the bicyclic molecules. Similar  profile shapes were observed in monocyclic molecules.\cite{hua_theoretical_2020, wei_vibronic_2022} Interpretation of major stick transitions of each isomer indicated that only a few (1--3), low-frequency (288.6--1507.4 cm$^{-1}$), nearly in-plane, ring-deformation modes are active [Figs. S1(c), S11(c)-S15(c)]. 

\subsection{Effects of  mode mixing}
\begin{figure*}
\includegraphics[width=1.0\textwidth]{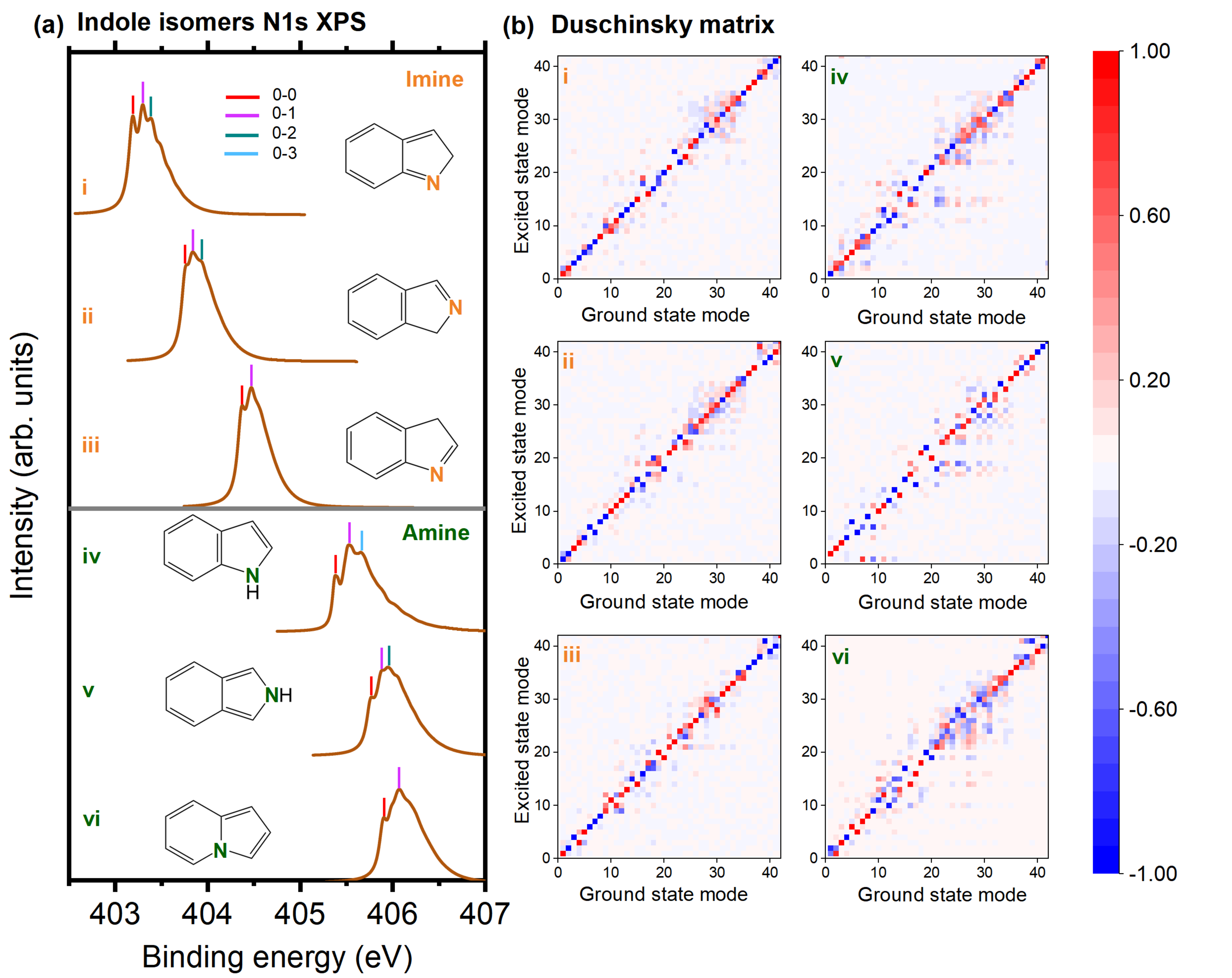}
\caption{(a) Comparison of simulated vibrationally-resolved N1s XPS spectra of six indole isomers: 2\textit{H}-indole (i), 1\textit{H}-isoindole (ii),  3\textit{H}-indole (iii), indole (iv),  2\textit{H}-isoindole (v), and  indolizine (iv). i-iii and iv-vi contain an imine and an amine N, respectively. (b) The  Duschinsky matrix of each isomer. The elements are generally more ``diffused'' off the diagonals  for amine (right) than imine (left) nitrogens. 
}
\label{1n}
\end{figure*}

The Duschinsky rotation matrix $\mathbf{J}$  carries information on the strengths of mode mixing.  The matrix elements of each indole isomer are visualized in Fig. \ref{1n}(b). Note that the range of each entry $J_{ij}$ is from -1 to 1. The absolute value of the element ($|J_{ij}|$) that is close to 0 and 1  mean respectively weak or strong mixing effect for the corresponding mode pairs $i$ and $j$). It is noted that in the literature,  absolute\cite{hebestreit_structures_2019, henrichs_excited_2021, peng_vibration_2010} or squared\cite{biczysko_first_2009} values are also often used (thus gives a range from 0 to 1). The three presentation methods are consistent, which will not influence our discussion.

As visualized in Fig. \ref{1n}(b), elements with the largest large absolute values are always at or near the diagonal. As we know, core ionization can lead to interchanges of the mode order. It is interesting to  find that for iv-vi (with amine N's) the corresponding matrix elements distribute farther off the diagonal than i-iii (with imine N's). This indicates a stronger mode mixing effect in amine than imine N's. In the three molecules iv-vi with amine nitrogens, the mode mixing happens between modes No. 25-35 (1178.5-1426.9 cm$^{-1}$), corresponding to nearly in-plane, ring-deformation vibrations. 
Analyses of the Duschinsky matrices for other molecules are provided in Figs. S18-S20, where the same conclusion for imine and amine nitrogens generally applies.
 
\subsection{Effects of CH$\leftrightarrow$N replacement}
\begin{figure*}
\includegraphics[width=0.45\textwidth]{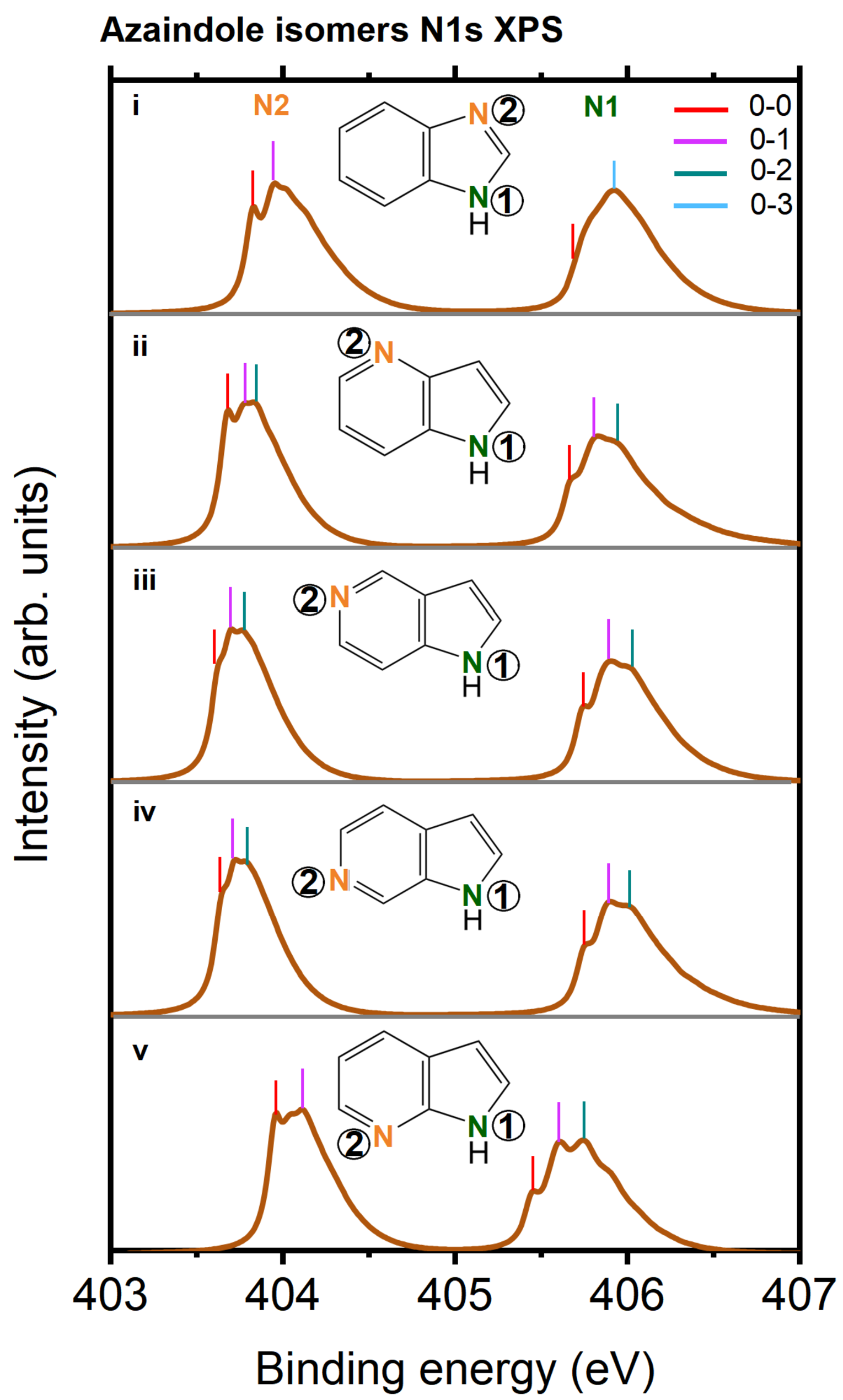}
\caption{Comparison of simulated vibrationally-resolved  N1s XPS spectra of five azaindole isomers. In each isomer, the amine nitrogen N1 is always fixed while  the imine nitrogen N2 is altered: benzimidazole (i), 4-azaindole (ii), 5-azaindole (iii),  6-azaindole (iv), and  7-azaindole (v). Peaks are interpreted.
}
\label{2nxps}
\end{figure*}

Figure \ref{2nxps} shows the spectra of five azaindole isomers, each with one amine nitrogen N1 and one imine nitrogen N2. N1 is always at a fixed position, while N2 replaces the -CH group at different sites. The effects of isomerization have been discussed above for indoles (Section \ref{sec:isomer:indole}).  Here the CH$\leftrightarrow$N replacement is a special kind of isomerization. Besides this effect, another theme here is to analyze the effects of N1-N2 interactions on the spectra, as discussed below (Section \ref{sec:nn:interaction}).


For each isomer, the two peaks from N1 and N2 are clearly separated. Responding to the position change, the N2 peak varies within an energy range of 0.3 eV between 403.8 and 404.1 eV, and its fine structure shows a clear isomer-dependent signature. Each molecule is identified to have 2-3 active vibrational modes. They are all in-plane modes within a frequency range of 246.5--1576.2 cm$^{-1}$ [Figs. S6(d)-S10(d)]. For $m$-azaindole ($m$=4, 5, 6, 7), all active modes of arise from N2.  Structurally, benzimidazole is the only molecule that contains both nitrogens in the five-membered pyrrole ring. It is interesting to find that the active modes of benzimidazole are ring deformation vibrations localized only on the five-membered ring [Fig. S6(d)]. For other isomers, the actives are more delocalized [Figs. S7(d)-S10(d)].

Besides, the FCH-optimized structure of benzimidazole is clearly distinguished from the other four isomers (Table \ref{tab:st}). The N1$^*$--C bond length of benzimidazole (1.55 {\AA}) is significantly greater than that of the four $m$-azaindole  isomers (1.45-1.47 {\AA}).  The same conclusion applies for N2: the N2$^*$--C distance in  benzimidazole is 1.39 {\AA}, while in the rest isomers, the length is 1.33--1.34 {\AA}.  Similarly, we also find that the bond angle  $\angle$C--N$_\text{a}^*$--C ($\angle$C-N$_\text{i}^*$--C)  for the amine N1 (imine N2) is 107.6--108.0$^\circ$ (121.4--124.6$^\circ$) for the four $m$-azaindole isomers. While for benzimidazole, the value is much smaller: 103.9 (108.3$^\circ$). Results indicated that the influence of CH$\leftrightarrow$N replacement within a five- or a six-membered ring is evidently different. 

\subsection{Effects of N-N interactions}\label{sec:nn:interaction}
Although N1 is always fixed in each azaindole isomer, it is interesting to find that the energy position of the N1 peak also varies evidently by 0.4 eV (from 405.8 to 406.2 eV), a value comparable to that of N2. The corresponding spectral profile of the N1 peak in each molecule also distinguishes well from each other. The change is attributed to the structural perturbation caused by N2. This reflects the interaction between the two nitrogens, which varies by molecules. The atom-atom interaction within a molecule was theoretically investigated and compared in p-, m-, o-aminophenols with para-, meta-, and ortho-positions of the -NH$_2$ and -OH groups by core excitations.\cite{hua_study_2016} Here the two nitrogens are connected via the delocalized $\pi$ orbital since alternating single and double bonds create a conjugated $\pi$ bond system across multiple atoms. 

From another point of view, the interaction between the two nitrogens can be read by comparing the spectra of 5- and 6-azaindole (iii and iv). In these two molecules, N1 and N2 have the largest spatial separation and consequently, the weakest interaction (similar to the role of p-aminophenol among all aminophenols\cite{hua_study_2016}), the resulting spectra (both the energies and profiles) are close to each other. 

\subsection{Limitation of the FCH-DR approach}
Another extreme for the nitrogen-nitrogen interaction is when they bond together (the strongest interaction).  When the two N atoms bond directly to one another in the five-membered ring, it gives another isomer, 1\textit{H}-indazole.  We tried  to simulate its spectrum with the same protocol, but the optimized FCH-state geometry deviates much from the GS one (a much larger N$^*$-N distance). As a result, the consequent Franck-Condon calculations failed to converge. The existence of the N$^*$-N motif seems to be a difficult case (though it does not always fail) for our DFT optimizations within the FCH approximation.

Critically, the same problem was also encountered in paper I\cite{wei_vibronic_2022} for 1,2,3-triazine and 1,2,3,4- and 1,2,3,5-tetrazine  as well as in Ref. \cite{du_theoretical_2022} for N$_5$H (these molecules were abandoned in the final publications due to the failure). The essential problem lies in locating the minima of the core-ionized isomer (\textbf{min FCH}), while a single-point (vertical) FCH  calculation at \textbf{min GS} usually works. This may be the limitation of the FCH-DR method when there is a tendency to significantly elongate the bond length at the ionization center. An approximate solution is to use the linear coupling model (LCM)\cite{macak_luo_lcm_2000} instead of the Duschinsky rotation, since the former calculation can avoid locating the structure of \textbf{min FCH}. For a more accurate solution within the DR framework,  one shall try high-level electronic structure methods for such difficult cases, such as the multiconfigurational methods.\cite{zhangyu_nonlinear_2016} 

\section{Summary and Conclusions}

In summary, we have computed the vibrationally-resolved XPS spectra for a family of 17 indole-based bicyclic molecules at the DFT level to investigate the structure-spectroscopy relation and to yield general vibronic coupling rules of this family.  Structural variations cover common side chain substitutions, isomerizations, CH$\leftrightarrow$N replacements. Vibronic coupling was considered by the Franck-Condon approach with the inclusion of the Duschinsky rotation effect. Good agreement with available experiments was achieved both in absolute binding energies and fine structures, which validated reliable analyses. Binding energies, major vibronic transitions, corresponding active vibrational modes, and core ionization-induced local geometrical changes at N$^*$ were analyzed for each molecule. Our study in this work provides reliable and complete spectral data for future experimental and theoretical references and machine learning studies, and the rules summarized are insightful for a better understanding of the basic physical concept of vibronic coupling in the XPS process. 

In this work, not only the clear separation of binding energies of amine (larger) and imine (smaller) nitrogens, as reported in several early literatures,\cite{du_theoretical_2022, zhang_accurate_2019} were confirmed, but more importantly, it is interesting to find the same separation rules also for vibronic properties. (1) The $\Delta {\varepsilon}_{0}$ value (zero-point vibrational energy of the FCH  state as compared to the ground state) is positive for all amine N's and negative for all imine N's. This indicates that the change in the curvature of the excited-state PES induced by core ionization depends on the local structure at the N$^*$ center, which can be steeper (for amine N's) or smoother (for imine N's). (2) Amine nitrogens exhibit stronger mode mixing effects than imine ones. (3) 1s ionization at amine N's always leads to an elongation of N$^*$-C and shortening of the N$^*$-H bond lengths; while for imine N's, the structural changes in N$^*$-C are much smaller and both elongation and shortening exist. (4)  Ionization at amine (imine) nitrogens always leads to a decrease (increase) in the bond angle $\angle$C-N$^*$-C. (5) 1s ionization in an amine N leads to a larger global structural change (indicated by the RMSD between \textbf{min GS} and \textbf{min FCH}) than an imine N.

\section{Acknowledgments}
Financial support from the National Natural Science Foundation of China (Grant No. 12274229) and the Postgraduate Research \& Practice Innovation Program of Jiangsu Province (Grant Nos. KYCX22\_0425) is greatly acknowledged.

\providecommand{\newblock}{}

\begin{table*}
    \centering
    \caption{
Vertical and adiabatic ionization potentials ($I^\text{vert}$ and $I^\text{ad}$), 0-0 transition energies ($E^\text{00}_\text{DR}$), and $\Delta {\varepsilon}_{0}$ of all 17 molecules simulated by B3LYP. Deviations to corresponding experiments are included in parentheses. All energies are in eV. See molecular structures and N indexes (omitted for mono-nitrogen molecules) in Fig. \ref{fenzi}(a).
} \label{tab:ip}
\begin{threeparttable}
    \begin{tabular}{rclllccr}

        Molecule & N$^*$ & Expt. & $I^\text{vert}$ & $I^\text{ad}$ &$E^\text{DR}_\text{00}$ & $\Delta {\varepsilon}_{0}$  \\ \hline
        Indole & ~ & 405.82\tnote{a} & 405.68 (-0.14) & 405.29 (-0.53) & 404.89 & -0.02  \\ 
        3-Methylindole & ~ & 405.70\tnote{b} & 405.46 (-0.24) & 405.11 (-0.59) & 405.01 & -0.01  \\ 
        3-Formylindole & ~ & 406.36\tnote{a} & 406.24 (-0.12) & 405.90 (-0.46) & 405.86 & -0.05  \\ 
        Adenine &N1 &404.4\tnote{c}& 403.91(-0.09) & 403.62 (-0.78)& 403.97 & +0.04  \\ 
        ~&N2 & 404.4\tnote{c}  & 404.04 (-0.40) & 403.74 (-0.66) & 403.33 & +0.04   \\ 
        ~&N3 &404.4\tnote{c} & 404.45 (+0.05) & 404.19 (-0.21) & 404.22 & +0.03   \\ 
        ~&  N4 & 406.7\tnote{c} & 406.37 (-0.33) & 405.87 (-0.83) & 405.79 & -0.08  \\
        ~& N5  & 405.7\tnote{c} & 405.47 (-0.33) & 405.10 (-0.60) & 405.06 & -0.04  \\         
        Purine &N1 & ~ & 404.48 & 404.27 & 404.29 & +0.01  \\ 
                ~&N2 & ~ & 404.83 & 404.60 & 404.60 & +0.00  \\ 
                ~&N3 & ~ & 404.73 & 404.49 & 404.51 & +0.02  \\ 
                ~&  N4 & ~ & 406.75 & 406.28 & 406.18 & -0.10  \\         
         Benzimidazole & N1 & ~ & 406.17 & 405.76 & 405.68 & -0.08  \\ 
        ~&N2 & ~ & 404.00 & 403.78 & 403.82 & +0.04  \\ 
        4-Azaindole & N1 & ~ & 405.98 & 405.68 & 405.65 & -0.03  \\ 
        ~&N2 & ~ & 403.81 & 403.63 & 403.67 & +0.04   \\ 
        5-Azaindole & N1 & ~ & 406.07 & 405.77 & 405.33 & -0.04  \\ 
        ~&N2 & ~ & 403.76 & 403.57 & 403.62 & +0.05  \\
        6-Azaindole & N1 & ~ & 406.07 & 405.77 & 405.73 & -0.04  \\ 
        ~&N2 &~& 403.77 & 403.60 & 403.64 & +0.04    \\ 
        7-Azaindole & N1 & ~ & 405.82 & 405.48 & 405.44 & -0.04  \\ 
                    ~&N2 & ~ & 404.10 & 403.92 & 403.95 & +0.03   \\ 
        1\textit{H}-isoindole & ~ & ~ & 403.88 & 403.71 & 403.75 & +0.04  \\        
        2\textit{H}-indole & ~ & ~ & 403.31 & 403.13 & 403.18 & +0.05  \\ 
        3\textit{H}-indole & ~ & ~ & 404.52 & 404.36 & 404.36 & +0.00  \\ 
        2\textit{H}-isoindole & ~ & ~ & 406.07 & 405.80 & 404.76 & -0.04  \\ 
        Indolizine & ~ & ~ & 406.18 & 405.92 & 404.99 & -0.03 \\ 
        7-Azaindazole & N1 & ~ & 406.43 & 405.97 & 405.90 & -0.07   \\
        ~&N2 &~& 405.19 & 405.03 & 405.03 & +0.00    \\ 
       ~& N3 &~&404.51 &  404.31 & 404.33 & +0.02    \\ 
        Pyrazolo[1,5-a]pyrimidine & N1 & ~ &407.15 & 406.75 & 406.68 & -0.07   \\
        ~& N2 &~& 404.66 & 404.45 & 404.66 & +0.01  \\ 
        ~&N3 &~&  404.80 &  404.60 & 404.62 & +0.02     \\ 

    \end{tabular}
    \begin{tablenotes}
\item[a] Plekan et al.\cite{plekan_experimental_2020}
\item[b] Zhang et al.\cite{zhang_electronic_2009}
\item[c] Vall-llosera et al.\cite{plekan_theoretical_2008}
\end{tablenotes}   
\end{threeparttable}
\end{table*}

\begin{table*}
\centering
    \caption{Bond lengths (in {\AA}) and angles (in $^\circ$) at the ionization center N$^*$ [amine nitrogen N$_\text{a}^*$ or imine nitrogen N$_\text{i}^*$; see definitions in Fig. \ref{fenzi}(b,c)]  at its optimized FCH-state  geometry (\textbf{min FCH}). Structural changes with respect to the ground state (\textbf{min GS}) are included in parenthesis.  RMSD values (in \AA) of the two superimposed geometries are also given. See molecular structures as well as definitions of bond angles and N indexes (omitted for mono-nitrogen molecules) in Fig. \ref{fenzi}(a).
    }    \label{tab:st}
\resizebox{\textwidth}{!}{
\begin{threeparttable}

    \begin{tabular}{cccccccc}
    \hline\hline 
        Molecule &N$^*$& N$_\text{a}^*$-C& N$_\text{a}^*$-H & N$_\text{i}^*$-C &  $\angle$C-N$_\text{a}^*$-C&$\angle$C-N$_\text{i}^*$-C& RMSD \\  \hline
        Indole & & 1.45 (+0.07); 1.46 (+0.08) & 0.98 (-0.03)  &  ~ &107.8 (-1.5)  & ~ &  0.10   \\
       3-Methylindole & & 1.45 (+0.07); 1.47 (+0.09)   &0.98 (-0.03)  &  ~ &107.3 (-1.8)  &  ~ & 0.09   \\
       3-Formylindole && 1.45 (+0.09); 1.46 (+0.07)   & 0.98 (-0.03)  &  ~ & 108.0 (-1.6)  &~ &  0.07   \\
      Adenine & N1 & ~ & & 1.34 (-0.00); 1.37 (+0.03) & ~& 124.0 (+5.3) &0.05   \\ 
        ~ & N2 & ~ &  & 1.33 (-0.01); 1.36 (+0.03)  &  & 115.6 (+4.2) & 0.04  \\ 
        ~ & N3 & ~ &  &  1.31 (+0.00); 1.38 (+0.00) &  & 107.7 (+3.5) &0.04 \\ 
        ~ & N4 & 1.44 (+0.06); 1.53 (+0.15)&  0.98 (-0.03) & ~ & 104.2 (-2.6)&  & 0.06  \\ 
        ~ & N5 & 1.48 (+0.13)&  0.98 (-0.02) &  & ~ & ~ & 0.06  \\ 
     Purine & N1 &   & &  1.33 (-0.00); 1.33 (-0.01) &  & 119.7 (+1.3) & 0.04  \\ 
        ~ & N2 &   & & 1.31 (-0.01); 1.37 (+0.03)  &  & 117.0 (+4.6) &0.03  \\ 
        ~ & N3  &   & &  1.31 (+0.01); 1.38 (-0.00) &  &107.8 (+3.5) & 0.03  \\ 
        ~ & N4 & 1.43 (+0.06); 1.56 (+0.18)  &  0.98 (-0.03) & ~  & 103.2 (-3.1) &  &0.05   \\   
        Benzimidazole &N1& 1.43 (+0.05); 1.55 (+0.18)  & 0.98 (-0.03)  & ~  & 103.9 (-3.0)  &~ &  0.06   \\ 
        ~&N2 & ~ & ~  &  1.31 (+0.00); 1.39 (-0.00)  &  ~ &108.3 (+3.4)  & 0.03   \\ 
        4-Azaindole &N1& 1.44 (+0.07); 1.47 (+0.09)  & 0.98 (-0.03)  & ~ &107.6 (-1.4)  & ~ & 0.07   \\ 
        ~ &N2& ~ & ~ &  1.33 (-0.01); 1.34 (+0.01)  &   ~ &121.4 (+5.3)  & 0.03   \\ 
        5-Azaindole &N1& 1.44 (+0.07); 1.47 (+0.08)  &  0.98 (-0.03)  & ~ &108.0 (-1.2)  & ~ &  0.05   \\ 
        ~ &N2& ~ & ~ &  1.33 (+0.00); 1.34 (-0.01)    &  ~ & 124.6 (+6.1)  &0.04   \\ 
        6-Azaindole &N1& 1.45 (+0.07); 1.46 (+0.08)    & 0.98 (-0.03)  & ~ & 107.6 (-1.3)  & ~ & 0.07   \\ 
        ~ &N2& ~ &   &1.33 (+0.00); 1.35 (-0.00)   &  ~ &124.4 (+5.8)  & 0.04   \\ 
        7-Azaindole &N1& 1.45 (+0.07); 1.46 (+0.08)   &  0.98 (-0.03)  & ~ &  107.7 (-1.3)  &~ & 0.03   \\ 
        ~ &N2& ~ & ~  &  1.31 (-0.01); 1.35 (+0.02)  &  ~ &119.9 (+5.3)  & 0.03   \\   
       1\textit{H}-isoindole &&  &   &  1.27 (-0.01); 1.42 (-0.00)  &  & 111.0 (+4.2) & 0.04   \\
       2\textit{H}-indole &&  &  &    1.32 (-0.01); 1.35 (+0.02) &    &119.9 (+5.3) & 0.03   \\ 
       3\textit{H}-indole &&  &  & 1.32 (+0.02); 1.43 (-0.02)  &   &109.4 (+3.8) & 0.03   \\ 
       2\textit{H}-isoindole &&1.43 (+0.07); 1.43 (+0.07)  &  0.98 (-0.03) &   &  110.1 (-1.9) &   & 0.03   \\
      Indolizine & &  1.42 (+0.05); 1.43 (+0.06); 1.47 (+0.05) &  &  &  108.1 (-0.8) &   & 0.03   \\
     7-Azaindazole &N1 &    1.41 (+0.05); 1.54 (+0.18)\tnote{a}  & 0.98 (-0.02) &  &  109.6 (-2.6)\tnote{c} &   & 0.05   \\
     &N2 &  & &  1.33 (+0.01); 1.35 (-0.01)&   &  108.9 (+2.2)\tnote{d} & 0.03   \\
      &N3 &  & & 1.32 (-0.01); 1.35 (+0.02)\tnote{b}&   &  119.5 (+5.4) & 0.03   \\
    Pyrazolo[1,5-a]pyrimidine &N1 &  1.39 (+0.04); 1.47 (+0.06); 1.49 (+0.14)\tnote{a}  &  & &  110.5 (-2.2)\tnote{c} &   & 0.05   \\
     &N2 &  & & 1.33 (+0.02); 1.34 (-0.01)&   &  122.2 (+5.0) & 0.03   \\
      &N3 & & & 1.35 (+0.02)\tnote{b}; 1.37 (+0.03) &   &  106.3 (+2.5)\tnote{d} & 0.03   \\
   \hline\hline   \end{tabular}
        
\begin{tablenotes}
\item[a] N$_\text{a}^*$-N.
\item[b] N$_\text{i}^*$-N.
\item[c] $\angle$C-N$_\text{a}^*$-N.
\item[d] $\angle$C-N$_\text{i}^*$-N.
\end{tablenotes}
\end{threeparttable}
}
\end{table*}

\end{document}